\documentclass[final,journal]{IEEEtran}

\usepackage{graphicx,cite}

\newcommand{\N}{\textsf{N}}

\begin{document}

\title{
A Constrained Coding Approach to Error-Free Half-Duplex Relay
Networks\thanks{\noindent Submitted for publication on December 13, 2011;
revised November 17, 2012 and May 28, 2013.\newline\indent F. R.
Kschischang is with the Electrical and Computer Engineering Department,
University of Toronto, 10 King's College Road, Toronto, Ontario M5S 3G4,
Canada  (e-mail: frank@comm.utoronto.ca), Tobias Lutz is with the Lehrstuhl
f\"ur Nachrichtentechnik, Technische Universit\"at M\"unchen, D-80290
M\"unchen, Germany (e-mail: tobi.lutz@tum.de).  The work of Frank R.
Kschischang was supported by a Hans Fischer Senior Fellowship from the
Technische Universit\"at M\"unchen, Institute for Advanced Study, funded by
the German Excellence Initiative.
}}

\author{Frank R. Kschischang \IEEEmembership{Fellow, IEEE} and Tobias Lutz} 

\IEEEpubid{0000--0000/00\$00.00~\copyright~2012 IEEE}

\maketitle

\begin{abstract}
\ifCLASSOPTIONonecolumn\relax\else\boldmath\fi
We show that the broadcast capacity of an infinite-depth tree-structured
network of error-free half-duplex-constrained relays can be achieved using
constrained coding at the source and symbol forwarding at the relays.
\end{abstract}


\section{Introduction}

\begin{figure}[tbp]
  \centering
  \includegraphics[scale=0.6667]{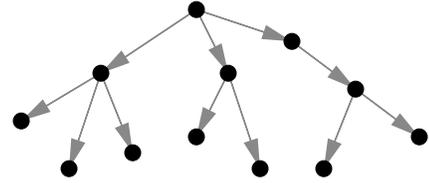}
  \caption{A broadcast tree of depth $D=3$.}
  \label{fig:tree}
\end{figure}

\IEEEPARstart{I}{nformation}
transmission through a relay channel or network with error-free and/or
half-duplex-constrained relays is a problem that has been considered by
several authors
\cite{VanMeu92,Kra04,Kra07,VijWonLok07,LuKrHa10,PoSi10,LuHaKo12}.  In this
paper the focus is on directed trees of error-free half-duplex-constrained
relays, as shown in Fig.~\ref{fig:tree}.  Such networks include a chain of
relays as a special case.  The transmission objective is to broadcast
information from a source (situated at the root of the tree) to all network
nodes, each of which is half-duplex constrained.  In each time slot, a node
either receives (without error) the transmission of its parent, or
broadcasts information to its children, but it may not do both.

More precisely, we assume that transmission between nodes in the network
occurs in discrete time-slots.  Let $\mathcal{Q}:=\{0,\dots,q-1\}$ be a
$q$-ary transmission alphabet, let $\N$ be an additional symbol which
indicates a channel use without transmission, and let
$\mathcal{X}:=\mathcal{Q}\cup\{\N\}$.  In any given time-slot, each node of
the network broadcasts a symbol $x \in \mathcal{X}$ to its children;  the
node is said to be {\sc on} if $x \in \mathcal{Q}$; otherwise $x = \N$ and
the node is said to be {\sc off}.

The half-duplex constraint is captured as follows.  When a relay is {\sc
off}, it is connected to its parent through a noiseless $(q+1)$-ary channel
with alphabet $\mathcal{X}$, and so receives the transmission from its
parent without error.  When a relay is {\sc on}, it cannot receive, so the
symbol sent by its parent is erased.

The simplest approach to information broadcasting is to require each
network node to be {\sc off} half of the time, organized in deterministic
fashion so that a node is {\sc off} whenever its parent might be {\sc on}.
Nodes simply forward what they receive, resulting in a transmission rate of
$0.5 \log_2(q+1)$~bits per symbol (b/sym).  The broadcast capacity, on the
other hand, approaches \cite{LuHaKo12}
\begin{equation}
C(q) := \log_2 \left( \frac{ 1 + \sqrt{4q+1} }{2} \right)~\textrm{b/sym}
\label{eqn:Cq}
\end{equation}
as the tree-depth becomes large.  In the binary case, deterministic
store-and-forward achieves $0.5$~b/sym whereas $C(1) = \log_2\phi=0.6924$,
where $\phi$ is the golden ratio.  For trees of finite depth, even greater
rates are possible.  For example, for trees of depth $D=2$, a rate of
$0.7729$~b/sym is achievable in the binary case.  It is clear that
deterministic store-and-forward falls short of the maximum possible
transmission rate.  However, to achieve the broadcast capacity of trees of
finite depth requires a sophisticated coding approach based on coding
additional information in the {\sc on}-{\sc off} patterns of the nodes
\cite{LuHaKo12}.

We note that {\sc on}-{\sc off} patterns have also been exploited for
neighbor discovery in half-duplex-constrained networks using a compressed
sensing approach \cite{ZhLuGu10,GuZh10,Applebaum_ChoirCodesAllerton}.
Another problem, namely a line of three nodes where the first two nodes are
half-duplex sources and where all nodes are connected by packet erasure
channels, was addressed in~\cite{LuKl11} within a queuing-theoretic
framework.  In~\cite{ZhGu10} a Gaussian point-to-point channel with a
sender subject to a duty cycle constraint (e.g., a half-duplex constraint)
and an average power constraint is considered.  Interestingly, the optimal
input distribution is shown to be discrete, i.e., a modulated {\sc on}-{\sc
off} signaling scheme is capacity-achieving.

In this paper we will present a broadcasting scheme, based on constrained
coding, that preserves the simplicity of the store-and-forward approach,
but achieves a higher transmission rate than deterministic
store-and-forward.  In particular, we show that we can achieve a broadcast
rate of $C(q)$ in any error-free half-duplex-constrained tree network using
constrained coding at the source and symbol forwarding at the relays.

\IEEEpubidadjcol

\section{Constrained Coding Background} 

The approach we take to broadcasting in a tree of half-duplex-constrained
nodes uses tools from constrained coding (or symbolic dynamics); see, e.g.,
\cite{LindMarcus95,MaRoSi01}.  In a nutshell, the field of constrained
coding studies mappings from unconstrained input sequences to output
sequences obeying certain constraints.  The constraints are often expressed
by specifying forbidden sub-blocks, i.e., subsequences that are not
permitted to occur in any output sequence.  A classical example is the
\textit{golden mean shift}, which is the set of binary sequences in which
the sub-block $11$ never occurs.  Constrained coding has found many
applications in magnetic and optical recording systems.

The \emph{capacity} of a constrained system, which is the maximum rate at
which unconstrained binary data may be mapped to constrained output data,
is defined as
\[
C = \limsup_{n \rightarrow \infty} \frac{1}{n} \log_2 N(n) ~\textrm{b/sym},
\]
where $N(n)$ denotes the number of sequences in the output alphabet having
length $n$ and satisfying the given constraint.  For example, the golden
mean shift satisfies the Fibonacci recurrence: for $n \geq 2$,
\[
N(n) = N(n-1) + N(n-2), \textrm{~with~}N(0)=1,~N(1)=2.
\]
From this it can be shown that the golden mean shift has $C = \log_2 \phi$,
where $\phi = (1+\sqrt{5})/2$ is the golden ratio (a result that explains
the name ``golden mean shift'').  Interestingly, the golden ratio also
arises in the analysis of the trapdoor channel~\cite{PeHaCuRo08} with
feedback, where it is shown that the capacity equals $\log_2\phi$.

It is well known that the capacity of certain constrained systems can be
obtained via an irreducible, lossless graph presentation of the constraint
\cite{LindMarcus95}.  If $G$ is such a presentation, and $A_G$ is the
adjacency matrix of $G$, then
\[
C = \log_2 \lambda(A_G),
\]
where $\lambda$ is the largest of the absolute values of the eigenvalues of
$A_G$.  This formulation of capacity will be used in the sequel.

\section{Code Construction}

We now describe the constrained coding approach taken in this paper.  The
transmission protocol is trivial, amounting to simple symbol-forwarding:
during any given time-slot, every non-source node simply forwards (to all
of its children) the symbol it has received from its parent during the
previous time-slot.   Correct forwarding is achieved provided that nodes
obey the half-duplex constraint, i.e., that they are never {\sc on} when
their parent node might be {\sc on}.  Under the symbol-forwarding protocol,
this is accomplished if and only if the source is never itself {\sc on} in
two adjacent time-slots.

\begin{figure}[th]
\centerline{\includegraphics[scale=0.6667]{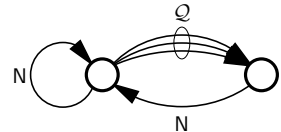}} \caption{Graph
presentation of half-duplex constraint under symbol-forwarding.}
\label{fig:constraint}
\end{figure}

Thus we arrive naturally at a constrained coding problem: the source may
emit any sequence of symbols drawn from $\mathcal{X}$ satisfying the
constraint that no two adjacent symbols are drawn from $\mathcal{Q}$.  In
the language of symbolic dynamics, every transmitted sequence is drawn from
the shift of finite type denoted as $\mathcal{X}_{\mathcal{Q}^2}$ having
forbidden sub-block set $\mathcal{Q}^2 := \mathcal{Q}\times\mathcal{Q}$.
An irreducible, lossless graph presentation of this shift is shown in
Fig.~\ref{fig:constraint}.  When $q=1$, this shift is equivalent to the
golden mean shift.

The adjacency matrix of this presentation is given, as a function of $q$,
as
\[
A(q) = \left[
\begin{array}{cc}
1 & q \\
1 & 0
\end{array}\right],
\]
which has characteristic polynomial
\[
p_q(\lambda) = \lambda^2 - \lambda - q.
\]
The eigenvalues of $A(q)$ (the roots of $p_q(\lambda)$) are given as
\[
\lambda = \frac{1}{2}\left(1 \pm \sqrt{ 1 + 4q }\right),
\]
and the constrained capacity (the logarithm of the largest eigenvalue) is
given as
\[
C(q) = \log_2 \left(\frac{1 + \sqrt{ 1 + 4q }}{2}\right).
\]
Remarkably---and this is the central result of this paper---the constrained
coding approach achieves the broadcast capacity of infinite-depth trees,
but without the necessity of designing sophisticated timing codes as in
\cite{LuHaKo12}.

\begin{figure}[th]
\centering
\includegraphics[scale=0.6667]{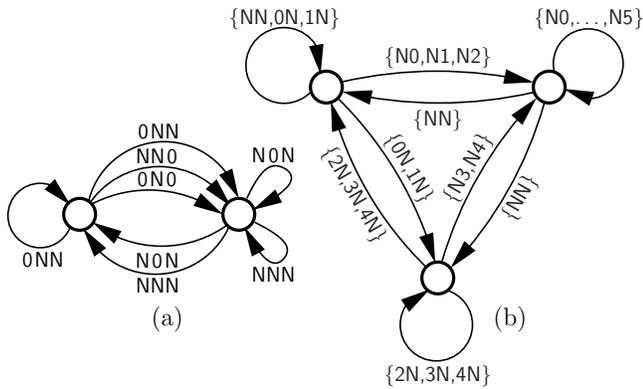}
\caption{Encoders for (a) $q=1$, $R=2/3$, $C_1=\log_2 \phi \approx 0.6942$
b/sym, (b) $q=6$,  $R=3/2$, $C_6 = \log_2 3 \approx 1.5850$ b/sym.}
\label{fig:golden}
\end{figure}

The capacity $C(q)$ can be approached using methods (e.g., the
state-splitting algorithm) from constrained coding.  Fig.~\ref{fig:golden}
provides two examples.  The first, in Fig.~\ref{fig:golden}(a), is a
standard example in constrained coding \cite{MaRoSi01} and gives a
rate-(2/3) encoder for $q=1$, which achieves more than $96$\% of the
capacity $C(1)=\log_2(\phi)$.  The second, in Fig.~\ref{fig:golden}(b), is
a rate-(3/2) encoder for $q=6$, which achieves more than $94$\% of the
capacity $C(6)=\log_2(3)$. The encoder can be constructed in three steps.
First, the second power graph is generated from the golden mean shift shown
in~Fig~\ref{fig:constraint}. Subsequently, the state with most outgoing
edges is split (according to the splitting criteria of the state-splitting
algorithm). In the final step, sufficiently many edges are deleted so that
eight outgoing edges per state remain.  Similar examples can readily be
constructed for other values of $q$.  For any given $q$, if the number of
encoder states is allowed to grow, $C(q)$ can be approached arbitrarily
closely.

\section{Discussion}

Table~\ref{table:achievable} compares, for $q=1$, rates achievable in
networks of finite depth $D$ using three approaches: the timing codes
presented in \cite{LuHaKo12}, the constrained coding approach of this
paper, and the deterministic store-and-forward approach.  The row labelled
$C$, which gives the maximum achievable rate (using timing codes), serves
as a benchmark for the other schemes.  We observe that $C(q)/C$, the
relative efficiency of constrained coding, rapidly converges to unity as
$D$ increases.  On the other hand, $0.5/C$, the relative efficiency of
deterministic store-and-forward, saturates at approximately $72\%$.  The
differences among the three approaches however become smaller as $q$
increases.

\begin{table}
\centering
\caption{Achievable Rates
in Networks of Finite Depth $D$ with $q=1$}
\label{table:achievable}
  \begin{tabular}{r|ccccc}
    $D$  & $2$ & $3$& $5$& $11$ &  $\infty$ \\ \hline\hline
    $C$ (b/sym) & $0.7729$ & $0.7324$&$0.7099$&$0.6981$&$0.6942$ \\ 
    $C(q)/C$ (\%)&$89.82 $&$94.79 $&$97.80 $&$99.44 $&$100 $ \\ 
    $0.5/C$ (\%)&$64.70 $&$68.27 $&$70.43 $&$71.62 $&$72.02 $ \\
  \end{tabular}
\end{table}

We observe that timing codes require nodes at different depths in a finite
tree to have different (carefully designed) {\sc on}-{\sc off} duty cycles
that depend on the depth of the tree.  It is shown in \cite{LuHaKo12} that
these duty cycles converge to a constant as the tree depth grows.  The
capacity $C(q)$, achieved both by timing codes and by the constrained
coding approach of this paper, is the maximum rate that can be achieved
with a constant {\sc on}-{\sc off} duty cycle throughout the network.

As a final remark, we note that it might be interesting to consider a
network model with noisy transmission links.  In this case, techniques that
combine constrained coding with error control (as in, e.g.,
\cite{Bl81,vWI01,CLU08}) may be helpful.


\begin{IEEEbiographynophoto}{Frank R. Kschischang}
received the B.A.Sc.\ degree (with honors) from the University of British
Columbia in 1985 and the M.A.Sc.\ and Ph.D.\ degrees from the University of
Toronto in 1988 and 1991, respectively, all in electrical engineering.  He
is a Professor and Canada Research Chair at the University of Toronto,
where he has been a faculty member since 1991.  Between 2011 and 2013 he
was a Hans Fischer Senior Fellow at the Institute for Advanced Study,
Technische Universit\"at  M\"unchen.

His research interests are focused primarily on the area of channel coding
techniques, applied to wireline, wireless and optical communication systems
and networks.  He is the recipient of the 2010 Killam Research Fellowship,
the 2010 Communications Society and Information Theory Society Joint Paper
Award and the 2012 Canadian Award in Telecommunications Research.  He is a
Fellow of IEEE, of the Engineering Institute of Canada, and of the Royal
Society of Canada.
  
During 1997-2000, he served as an Associate Editor for Coding Theory for
the \textsc{IEEE Transactions on Information Theory}.  He also served as
technical program co-chair for the 2004 IEEE International Symposium on
Information Theory (ISIT), Chicago, and as general co-chair for ISIT 2008,
Toronto.  He served as the 2010 President of the IEEE Information Theory
Society.
\end{IEEEbiographynophoto}

\vspace*{-2\baselineskip}
\begin{IEEEbiographynophoto}{Tobias Lutz}
was born in Krumbach, Germany, on May 2, 1980. He received the B.Sc.\ and
Dipl.-Ing.\ degrees in electrical engineering and information technology
from the Technische Universit\"at M\"unchen, Germany, in 2007 and 2008,
respectively.  Funded by a grant from the American European Engineering
Exchange program (AE3), he studied electrical and computer engineering at
the Rensselaer Polytechnic Institute, Troy, NY, during the academic year
2005/06.  Since 2008, he has been with the Institute for Communications
Engineering, Technische Universit\"at M\"unchen, where he is currently
working toward the Dr.-Ing.\ degree under the supervision of Prof.\ Gerhard
Kramer.  Since 2009 he has been studying financial mathematics at the
Ludwig Maximilian Universit\"at M\"unchen.  Tobias Lutz was awarded a
Qualcomm Innovation Fellowship in 2012. From August 2012 through March 2013
he was a visiting graduate student at Stanford University. His current
research interests include information and coding theory and their
application to wireless relay networks.
\end{IEEEbiographynophoto}
\end{document}